\documentclass[journal]{IEEEtran}

\usepackage{cite}
\usepackage{graphicx}
\usepackage{subcaption}

\makeatletter
\long\def\@makecaption#1#2{%
  \setbox\@tempboxa\hbox{{\fontsize{8}{9}\selectfont #1. #2}}%
  \ifdim \wd\@tempboxa >\hsize
    {\fontsize{8}{9}\selectfont #1. #2\par}%
  \else
    \hbox to\hsize{\hfil\box\@tempboxa\hfil}%
  \fi}
\makeatother

\usepackage{amsmath}
\usepackage{array}
\usepackage{xcolor}

\ifCLASSOPTIONcompsoc
 \usepackage[caption=false,font=normalsize,labelfont=sf,textfont=sf]{subfig}
\else
 \usepackage[caption=false,font=footnotesize]{subfig}
\fi

\usepackage{url}

\begin{document}

\title{Plug-and-Play Stability Certificates Compliant with Black-Box Models of Devices}
%
%
%
\author{Petr~Vorobev,~\IEEEmembership{Member,~IEEE,},~Andrey~Gorbunov,~\IEEEmembership{Member,~IEEE,},~Youhong~Chen,~\IEEEmembership{Member,~IEEE,}~Janusz~Bialek,~\IEEEmembership{Fellow,~IEEE}, and Balarko Chaudhuri,~\IEEEmembership{Fellow,~IEEE}
\thanks{P. Vorobev is with the School of Electrical and Electronic Engineering, Nanyang Technological University. E-mail: petr.vorobev@ntu.edu.sg}
\thanks{A. Gorbunov is with the School of Electrical and Computer Engineering, the University of Sydney. E-mail: andrey.gorbunov@sydney.edu.au.}
\thanks{Y. Chen is with the School of Electronic and Electrical Engineering, the University of Bath. E-mail: yc2930@bath.ac.uk}
\thanks{J. Bialek and B. Chaudhuri are with the Department of Electrical and Electronic Engineering, Imperial College London. E-mail: j.bialek@imperial.ac.uk, b.chaudhuri@imperial.ac.uk}
 }

\markboth{IEEE Transactions on Power Systems}%
{Shell \MakeLowercase{\textit{et al.}}: Bare Demo of IEEEtran.cls for IEEE Journals}

\maketitle

\begin{abstract}
This paper presents a new approach to the derivation of decentralized plug-and-play stability certificates that can be calculated using black-box admittance spectra, without requiring white-box models of devices. The method is based on analysing the properties of the nodal admittance matrix. The certificates ensure that the devices can be connected at any node of a network of any topology as long as the $R/X$ ratio is within a specified range. The approach is related to the \emph{passivity} concept, but uses frequency-dependent transformation matrices to check if a device, which is non-passive in a certain frequency range, can still satisfy a plug-and-play stability criterion. We illustrate the method using Grid Forming Inverters (GFMs) connected to IEEE $39$ node test-case.
\end{abstract}


\IEEEpeerreviewmaketitle

\section{Introduction}

The development of renewable energy and energy storage technologies in the recent years has led to unprecedented increase in the total number of Inverter-Based Resources (IBRs) connected to power grids around the world. There is a growing body of evidence that a high penetration of IBRs may cause multi-frequency oscillatory interactions in power systems, spanning sub- and super-synchronous regimes, driven by control–network coupling effects \cite{chen2026sub}. Unlike conventional generators, IBRs have non-trivial dynamics across a wide range of time-scales, which calls for revision of traditional modeling approaches that were developed for power grids. Moreover, as individual IBRs have much smaller capacity than conventional generators, their number in a network is significantly higher. For instance, the number of wind turbines connected on a transmission level to power grid in Great Britain is already close to $10,000$ units, compared to barely few hundred conventional synchronous machines for the same grid.

Complex dynamics of IBRs combined with a high number of individual units make direct modeling of power grid dynamics difficult due to high computational costs. Additionally, due to proprietary concerns, vendors do not usually disclose to the System Operator explicit white-box models of IBRs. Hence, any stability analysis method should ideally use only so-called black-box models of devices, i.e. the admittance spectrum obtained from a compiled EMT simulation code disclosed by the vendor or physical testing of a device. All of that means that conceptually new approaches to grid stability assessment have to be developed.

One of the promising approach, first proposed in relation to microgrids \cite{lasseter2002microgrids}, is the so-called, \emph{plug-and-play} concept where power grid components are individually tuned in a certain specific way to make them compatible with each other under a wide range of grid configurations. This concept is closely related to \emph{decentralized stability certification}: stability of the whole network should follow from local properties of its components, even as topology, operating conditions, and controller implementations change. A natural starting point for such approach is the impedance-based framework, in which stability is assessed by terminal admittances of devices and their interaction with the  network \cite{harnefors2007input}. However, this direct impedance/admittance-based method requires measuring the network impedance/admittance at the point of connection, which includes all existing devices and bound to a specific network configuration. Consequently, connecting every additional device necessitates repeating this measurement procedure, making such direct stability certification difficult to implement. Ideally, one would call for a method, that would not require the full reassessment of the grid properties for every configuration and device.

One way to realize plug-and-play stability certificates, is through the use of the \textit{passivity} concept \cite{willems1972dissipativeI} which states that any feedback or parallel interconnection of passive subsystems represents a passive system, therefore, it is stable. Some works have developed decentralized frameworks that certify stability from local passivity properties of bus dynamics \cite{spanias2020passivity}. Passivity has also been employed beyond the small-signal setting to establish decentralized transient-stability conditions for dispatchable virtual oscillator control (dVOC), based on a large-signal output-feedback passivity property with an explicit passivity index \cite{he2024passivity}. Other works have aimed to redesign converter controllers or augment them with filters, damping loops, or virtual impedance loops, so that the resulting terminal admittance becomes passive \cite{watson2021scalable,pates2019robust,gross2022compensating,chen2024unified}. This line of work provides a route toward plug-and-play stability, but at the cost of restrictive controller structures or implementation complexity. Moreover, practical controllers are frequently not passive over the whole frequency range, especially at low frequencies \cite{vorobev2019decentralized}, so strict passivity may exclude otherwise well-performing designs.

Recent works have therefore exploited generalizations of passivity concepts. This includes decentralized frequency-domain conditions for inverter-based microgrids that combine a bus-level passivity test with a second local quadratic-constraint condition on subsystems formed by each bus and its incident $RL$ lines \cite{laib2020decentralized}, decentralized criteria for converter-interfaced generation that use admittance at high frequencies and a transformed low-frequency input--output description based on active/reactive power and derivatives of polar voltage components, together with a slow--fast separation of the controller dynamics \cite{dey2022passivity}, and parametric decentralized certificates for grid-forming converter control obtained from dynamic loop-shifting and generalized frequency-domain inequalities \cite{haberle2025decentralized}. Related developments have also introduced weighted or extended passivity conditions for MIMO voltage-source inverter models \cite{chen2024extended}, as well as passivity-index analysis and passivity-oriented impedance shaping for converter--grid interactions \cite{wang2014stability,wang2018harmonic}, and decentralized gain--phase conditions \cite{huang2024gain, niehues2025small}. These works move beyond strict passivity and reduce the conservativeness of standard passivity tests. However, they retain the classical viewpoint in which stability is analysed through the interaction between a device and a suitably modelled network. As a consequence, the analysis is often either limited to a single point of common coupling \cite{chen2024extended} or dependent on explicit knowledge of the network \cite{huang2024gain}. Even when such network knowledge is available, additional simplifying assumptions are typically required, such as lossless lines \cite{siahaan2024decentralized} or a fixed $R/X$ ratio \cite{haberle2025decentralized, niehues2025small}, together with controller simplifications, e.g.\ neglecting internal voltage-current control loops \cite{haberle2025decentralized} or assuming exact $V$--$q$ droop relations \cite{niehues2025small}. Other approaches rely instead on structural assumptions on the controllers, such as explicit low-/high-frequency separation  \cite{dey2022passivity}.

In contrast, the present work derives plug-and-play conditions directly from the \emph{properties of the nodal admittance matrix itself}, without recasting the problem as a feedback interconnection between a device and a known external network. This leads to two main advantages. First, the approach is model-agnostic: it does not rely on a particular internal control structure, reduced-order approximation, or some simplification of control loops for power electronics devices (such as neglecting current or voltage control loops, etc.). Therefore our method can be applied to a broad class of devices through their terminal admittance alone. Second, it is \emph{black-box compliant}: because the conditions are expressed directly in the frequency domain, they can in principle be checked using measured or identified admittance spectra, rather than analytically derived models. In this way, the method keeps the compositional spirit of passivity while avoiding controller redesign, restrictive network assumptions, and the need to partition the system into a device and a surrounding grid. It also extends the generalized-passivity idea behind the decentralized multiplier construction of \cite{vorobev2019decentralized} into a systematic framework for arbitrary interconnections of certified components. The method is generally applicable to any devices but in this paper we concentrate on Grid Forming Inverters (GFMs).

The main original contributions of this paper are as follows:
\begin{enumerate}
\item We present a method that allows for formulation of model-agnostic plug-and-play stability certificates. The method is based on analyzing the properties of the nodal admittance matrix itself, without recasting the problem as a feedback interconnection between a device and a known external network

\item We show that the method can be equally well used for white- and black-box models, which is important for application to real-life power grids, where exact models of the connected devices are usually unavailable.

\item The stability certificate (if it exists) ensures that a device can be connected at any node of a network with any topology as long as the $R/X$ line ratio is within a specified range. This makes it especially useful for vendors who want to be able to sell their devices to be connected to any network.

\end{enumerate}

We demonstrate the validity of the method using IEEE $39$ bus test-case populated with a number of GFMs with full realistic models. We perform extensive simulations over a big number of different configurations to validate the plug-and-play capability of the method. We also assess the conservativeness of the method and show that the conservativeness is rather modest for the realistic settings. Our approach is modular, scalable, and can be used to analyze the stability of large-scale systems with many devices.

\section{Problem Formulation}

\subsection{Network Admittance}

The proposed approach starts from representing power system components in terms of effective admittances. A well known relation between the nodal currents and nodal voltages in a power system can be written in the following form:
\begin{equation}\label{classical_Y}
    \bf{I} = \bf{Y_N} \bf{V}
\end{equation}
here $\bf{I}$ is the vector of nodal currents injections, $\bf{V}$ - vector of nodal voltages (both voltages and currents are complex phasors), while $\bf{Y_N}$ is the network admittance matrix (where the subscript $N$ refers to network) with non-diagonal elements $Y_{ij}$ being the mutual admittances between the corresponding nodes and diagonal elements $Y_{ii}$ being the sum of the admittances of all the lines terminating at the bus $i$. We neglect shunt capacitances of the lines.

Elements of the admittance matrix $Y_{ik}$ are essentially the negative inverse impedances of the corresponding lines - $Y_{ik} = -(R_{ik} + j X_{ik})^{-1}$. Such a representation does not take into account electromagnetic degrees of freedom, which, in most cases, is a valid approximation for traditional power grids. However, this approximation can become invalid for grids with high share of inverter-based resources, and equation \eqref{classical_Y} has to be generalized for this case. 

First, instead of the complex phasor representation for voltage and current we need to adapt a representation in a form of two-component vectors - d- and q-components:
\begin{equation}
    V_i = [V_{id}, V_{iq}]^T;\qquad I_i = [I_{id}, I_{iq}]^T
\end{equation}

This representation is equivalent to Cartesian representation for conventional power flow studies. Likewise, the admittance matrix $Y$ now becomes a block matrix withe the elements being $2$x$2$ matrices. In order to derive the explicit expressions for them, we write the line currents equations taking into account the electromagnetic phenomena explicitly. In d-q reference frame Kirschoff's voltage law for the line between nodes $i$ and $k$ is:
\begin{subequations}\label{current_dynamics}
\begin{align}
    L_{ik} \frac{dI_{d,ik}}{dt} = V_{d,i} - V_{d,k} -\omega_0 L_{ik} I_{q,k} + R_{ik} I_{d,k} \\
    L_{ik} \frac{dI_{q,ik}}{dt} = V_{q,i} - V_{q,k} +\omega_0 L_{ik} I_{d,k} + R_{ik} I_{q,k}
\end{align}
\end{subequations}

Since the relations between the nodal voltages and line currents is no longer algebraic, we first take the Laplace transform of equations. This leads to the following expressions for the elements of the admittance matrix:
\begin{equation}\label{Y_line}
Y_{ik} = 
\begin{bmatrix}
R_{ik} + s L_{ik} & -\omega_0 L_{ik} \\
\omega_0 L_{ik} & R_{ik} + s L_{ik}
\end{bmatrix}^{-1}
\end{equation}

Here $R_{ik}$ and $L_{ik}$ are resistance and inductance of the line between nodes $i$ and $k$, and $\omega_0$ is the frequency of the common reference frame, which we assume to be equal to the AC frequency of the grid at the steady state.

\subsection{Components Admittance}

Our next step is to formulate the voltage-current relation for power systems components - loads and generators. Mathematically, there is no difference between those two classes, for convenience we will refer to both of them as \emph{devices}. Representation \eqref{classical_Y} is possible due to natural linear voltage-current relations for the network. For generators and loads this linearity no longer holds in general case, and we have to linearize the system equations first. Let us assume that the system has an equilibrium steady state with some values of nodal voltages and line (and nodal) currents, which will be marked with a superscript $0$. Perturbation of voltage and current from the equilibrium can be written: 
\begin{equation}\label{voltage_linearization}
    V_{d,i}(t) = V^0_{d,i} + \delta v_{d,i}(t);\qquad V_{q,i}(t) = V^0_{q,i} + \delta v_{q,i}(t)
\end{equation}
\begin{equation}\label{current_linearization}
    I_{d,ik}(t) = I^0_{d,ik} + \delta i_{d,ik}(t);\qquad I_{q,ik}(t) = I^0_{q,i} + \delta i_{q,ik}(t)
\end{equation}
Here $\delta v_{d,i}$, $\delta v_{q,i}$, $\delta i_{d,ik}$, and $\delta i_{q,ik}$ are the small perturbation of the corresponding variables from their steady-state values. 

After linearization and switching to frequency domain (i.e., performing Laplace transform),  we can represent every device (i.e., generator or load) in terms of the corresponding admittance matrix, that links together the perturbation of voltage at the bus to which the device is connected to the perturbation of the current ($\delta i_{d,i}(s)$ and $\delta i_{q,i}(s)$) that the device draws from the grid (this is a standard sign convention for admittance matrices):
\begin{equation}\label{Y_device}
\begin{bmatrix}
\delta i_{d,i}(s) \\
\delta i_{q,i}(s)
\end{bmatrix}
=
\begin{bmatrix}
Y_{dd}(s) & Y_{dq}(s) \\
Y_{qd}(s) & Y_{qq}(s)
\end{bmatrix}
\begin{bmatrix}
\delta v_{d,i}(s) \\
\delta v_{q,i}(s)
\end{bmatrix}
\end{equation}
We note, that this effective admittance representation is possible for any type of device, including "passive" - constant impedance loads. The concept of effective small-signal impedance/admittance for different power system devices (including IBRs) has been around for some time, we can refer the reader to an excellent paper by L.Harnefors \cite{harnefors2007modeling}. 

One of the advantages of the use of impedance/admittance is that it can be either calculated if the model of a component in question is available, or just directly measured by a procedure of impedance scan. If the component model is available in state-space representation one needs to write it in the standard input-output form:
\begin{subequations}
\begin{equation}
    \dot{x} = Ax + Bu \\
\end{equation}
\begin{equation}
    y = Cx + Du
\end{equation}
\end{subequations}
with the input chosen as $u=[v_d, v_q]^T$ and the output as $y=[i_d, i_q]^T$. Then the admittance matrix can be found as:
\begin{equation}
Y(s) = C(sI-A)^{-1}B + D
\end{equation}
For most of the power systems components the explicit expressions for admittance matrix are very cumbersome. Thus, for a droop-controlled grid-forming inverter \cite{pogaku2007modeling} each component of the admittance matrix has a denominator of $13$-th order in Laplace domain.

Getting back to the network representation \eqref{classical_Y} we note, that the equations are linear, so that the same equations can be used for the small perturbation of nodal voltages and nodal currents injections with the same admittance matrix as in \eqref{classical_Y}:
\begin{equation}\label{classical_Y_small}
    \delta i(s) = \bf{Y_N}(s) \delta v(s)
\end{equation}
Here $\delta i$ is the full vector of the perturbations of the nodal currents injections, i.e., $[\delta i_{d,1}, \delta i_{q,1}, \delta i_{d,2} ... ]^T$ and  $\delta v$ is the full vector of the perturbations of the nodal voltages, i.e., $[\delta v_{d,1}, \delta v_{q,1}, \delta v_{d,2} ... ]^T$.

\subsection{System Representation}

Equations \eqref{Y_device} and \eqref{classical_Y_small} can be merged together, since the nodal currents in the left-hand side of each equation are just opposite of each other. In the absence of external nodal currents - no external disturbance of the grid, we come to the following equation for the nodal voltage perturbations:
\begin{equation}\label{delta_v_main}
    ( \bf{Y_N}(s) + \bf{Y_D}(s)) \delta v(s) = 0 
\end{equation}
Here $\bf{Y_D}(s)$ is a block-diagonal matrix with diagonal elements being represented by individual device admittance matrices from \eqref{Y_device} connected to corresponding nodes. In case no device is connected to a node, the corresponding elements are zeros. Let us introduce the following denotation for the full system matrix: $\bf{Y_F} = \bf{Y_N} + \bf{Y_D}$.

Equations \eqref{delta_v_main} have non-trivial solutions under the following  condition:
\begin{equation}\label{det_main}
    \textrm{Det} \left[ \bf{Y_F}(s) \right] = 0
\end{equation}
This equation is only satisfied for certain values of $s=s_i$, which coincide with the eigenvalues of the system state matrix, and correspond to the system eigen-modes. Time evolution of each mode can be described as:
\begin{equation}
    \delta v(t) = \delta v(0) e^{s_i t} 
\end{equation}
where $\delta v(0)$ is the initial value of the corresponding mode of the nodal voltage perturbation vector. 

For the system to be stable, all the eigen-values must have negative real part, i.e. $\textrm{Re}[s_i] < 0 $ for all $i$. Formally, stability of the system can be certified by finding all the roots of equation \eqref{det_main} and showing that they all have negative real part. Such an approach, although it can be convenient for certain cases, is fully equivalent to a direct eigen-value analysis of the system. In particular, one needs the full information about the system configuration, white-box models of the devices, control settings, and operating point in order to certify stability using the above approach. However, using the above representation it is possible to formulate stability conditions in a fully decentralized - the, so-called, \emph{plug-and-play} way, by exploiting the properties of the admittance matrix and performing a special mathematical procedure - homotopy - a virtual gradual transformation of the system configuration from a fictitious stable state to the actual
state. Hence, instead of checking the position of the eigenvalues, we are checking whether the eigenvalues have crossed the imaginary axis during homotopy - indicating instability - or whether they did not thus indicating stability. The main advantage of homotopy is that checking whether or not the eigenvalues have crossed the imaginary axis can be done in a decentralized way hence enabling a plug-and-play approach.

\section{Fully Decentralized Stability Certificates}

\subsection{Homotopy}

\begin{figure}[htbp]
    \centering
    \includegraphics[width=0.40\textwidth]{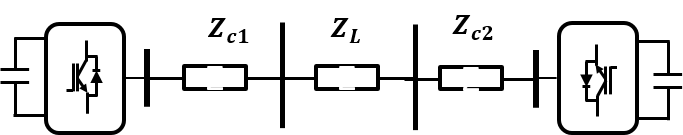}
    \caption{A test system of two grid-forming inverters connected by a line.}
    \label{fig:two_inv}
\end{figure}

We start homotopy by choosing some configuration for our system, that is definitely stable. This can be done by either gradually turning off all the controls, or by gradually transforming all the admittances of the system devices to a constant impedance form. Let us illustrate it by a simple example of a system of two grid-forming droop-controlled inverters \cite{pogaku2007modeling} connected by a line (Fig. \ref{fig:two_inv}). This simple example is used by us only for the purposes of illustrating the homotopy, not for actual stability studies. To start homotopy, we can always make the system stable by reducing the voltage and frequency droop gains to zero, so that all the roots of the equation \eqref{det_main} (which, as noted before, coincide with the system eigenvalues) for this system are located in the left-hand plane.  

\begin{figure}[htbp]
    \centering
    \includegraphics[width=0.35\textwidth]{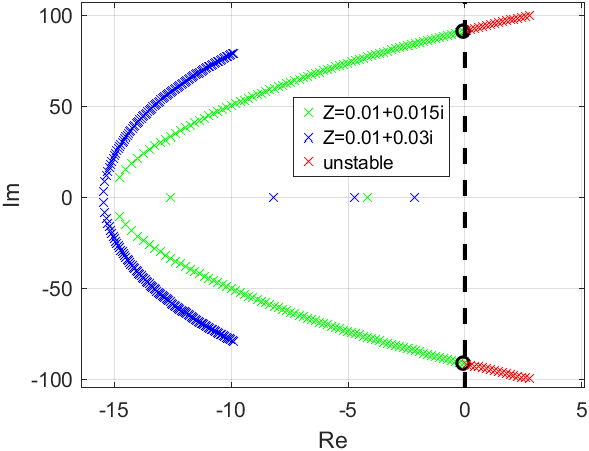}
    \caption{Trajectory of the eigenvalues for two systems with two-inverters during a homotopy. Eigenvalues that cross the imaginary axis are marked with “$\circ$".}
    
    \label{fig:eigs_zoom}
\end{figure}

Let us now gradually transform the system back to the initial configuration (i.e., increasing the voltage and frequency droop to their actual values) so that the roots of equation \eqref{det_main}  move in the complex plane. If our system is unstable, then at least one of the  root (i.e. a pole of the system) must necessarily cross the imaginary axis during the homotopy. We analyze two different systems with two inverters each, that differ by the size of the coupling impedances of the inverters. For both systems the droop gains (both voltage and frequency) are simultaneously varied from $0$ to $0.02$ pu. System No.1 has the coupling impedance of each inverter equal to $0.01 +j0.03$ pu, while system No.2 has the coupling impedance of $0.01 + j0.015$ pu. The trajectories of the roots of equation \eqref{det_main} for each system (only relevant modes are shown) during the homotopy are shown in Figure \ref{fig:eigs_zoom}: blue for System 1 and green for System 2. Clearly, system No.2 loses stability at some point during the homotopy - when two of the complex conjugate roots cross the imaginary axis. At the same time, system No.1 remains stable as none of its roots ever cross the imaginary axis.

The observation described in the previous paragraph, leads us to the following approach: if we can show that none of the roots of equation \eqref{det_main} cross the imaginary axis as we perform the homotopy \footnote{Here we assume that homotopy is regular enough, i.e. the poles and zeros of the system move continously along the homotopy path without zero-pole cancellations at $s=j\omega$, crossing from infinity, or sudden changes in the number of finite poles and zeros.} from a definitely stable configuration back to the initial configuration, then the system never loses stability, thus our initial system is stable. Since on the imaginary axis $s=j\omega$, we can state that the system is stable if:
\begin{equation}\label{det_freq}
   \forall \omega, \forall \alpha\in[0,1] \qquad \textrm{Det}\left[ \bf{Y_N}(j\omega,\alpha) + \bf{Y_D}(j\omega,\alpha)\right] \neq 0
\end{equation}
where we have introduced the homotopy parameter - $\alpha$ - with $\alpha=0$ corresponding to a starting, definitely stable, configuration, and $\alpha=1$ corresponding to the initial configuration.  In the rest of the paper, unless specifically indicated, all the equations involving admittance matrices are assumed to be satisfied for all values of frequency (from $-\infty$ to $+\infty$) and for values of the homotopy parameter $\alpha$ across the range $[0,1]$.

Equation \eqref{det_freq} does not yet give us any decisive advantage over solving equation \eqref{det_main} in terms of stability assessment. However, we can substitute condition \eqref{det_freq} with a more restrictive, yet much more useful condition of positive definiteness of a corresponding matrix. For this, let us recall that for any square matrix $A$ we have:
\begin{equation}
    \textrm{Det}(A) = \prod \sigma_n 
\end{equation}
where $\sigma_n$ is the $n$-th eigenvalue of the matrix $A$. I.e., the determinant of a matrix is a product of its eigenvalues. Thus, in order to show that a matrix determinant never turns to zero it is sufficient to show that none of its eigenvalues turn to zero. A useful special case of this is when a matrix is strictly positive definite - then all its eigenvalues are positive and, therefore none of them is zero. 

Returning to the matrix $\bf{Y_F} = \bf{Y_N} + \bf{Y_D}$ from \eqref{det_main} we note that its eigenvalues (not the same as eigenvalues of our system state matrix - the roots of equation \eqref{det_main}!) are in general complex, and the positive definiteness is not applicable to this matrix. However, if we take the Hermitian part of $\bf{Y_F}$ and demand its positive definiteness, it will be a sufficient (but not necessary) condition for $\bf{Y_F}$ to have no zero eigenvalues. Therefore, we can substitute the condition \eqref{det_freq} with the following one: 
\begin{equation}\label{pos_def_main}
    \bf{Y_F}(j\omega,\alpha) + \bf{Y_F}(j\omega,\alpha)^H  \succ 0
\end{equation}
which has to be valid for all values of frequency all the time as we perform the homotopy from the starting stable state to our state of interest. If this condition holds, then the system is stable. Here the superscript $H$ denotes Hermitian conjugation. Condition \eqref{pos_def_main} in general is more restrictive than condition \eqref{det_freq}, however, it provides a way to formulate the system stability conditions in a fully decentralized form. This is possible due to the following fact from linear algebra: if there are two matrices $\bf{A}$ and $\bf{B}$ and both $\bf{A}\succ 0$ and $\bf{B} \succ 0$, then $(\bf{A}+\bf{B})\succ 0$. We can use this property to our advantage to develop really decentralized stability conditions. We note here, that \emph{strict} positive definiteness is essential in \eqref{pos_def_main} - positive semi-definiteness is not sufficient. 

\subsection{Decentralized Conditions}

The above stated positive definiteness of the system admittance matrix (network plus devices) - equation \eqref{pos_def_main} - can be conveniently recast into a decentralized form using the properties of the admittance matrix. Let us start from the case when the network configuration is fixed and known, we later relax this assumption. Then, using the additivity property we can restate \eqref{pos_def_main} in the following way: 
\begin{equation}\label{pass_network}
    \bf{Y}_{N}(j\omega,\alpha) + \bf{Y}_{N}^H(j\omega,\alpha) \succ 0, 
\end{equation}
\begin{equation}\label{pass_devices}
    \bf{Y}_{D}(j\omega,\alpha) + \bf{Y}_{D}^H(j\omega,\alpha) \succ 0, 
\end{equation}
The latter equation we can split even further - down to the level of individual devices, since the devices admittance matrix $\bf{Y}_D$ is block-diagonal: 
\begin{equation}\label{pass_devs}
    \bf{Y}_{D,ii}(j\omega,\alpha) + \bf{Y}_{D,ii}^H(j\omega,\alpha) \succ 0, 
\end{equation}
Conditions \eqref{pass_network} and \eqref{pass_devices} resemble (but not exactly equivalent to) the classical \emph{passivity} concept - components with positive definite admittance matrices and poles in the left-hand plane are passive, and the whole system composed from such components is also passive, hence, stable.

Although conditions \eqref{pass_network} and \eqref{pass_devices} can be used in some cases to certify stability, we will not be using them since they are almost never satisfied for realistic models of devices. Instead, they serve us a methodological purpose to make the derivation of our main contribution clearer. In fact, condition \eqref{pass_network} is always satisfied for a network composed of RLC elements. Conditions \eqref{pass_devices} and/or \eqref{pass_devs}, however, are not automatically satisfied for most of the power grid components (except lines and/or constant impedance loads), so modifications to the method are needed if it is to be used for real-life systems.

\begin{figure}[htbp]
    \centering
    \includegraphics[width=0.40\textwidth]{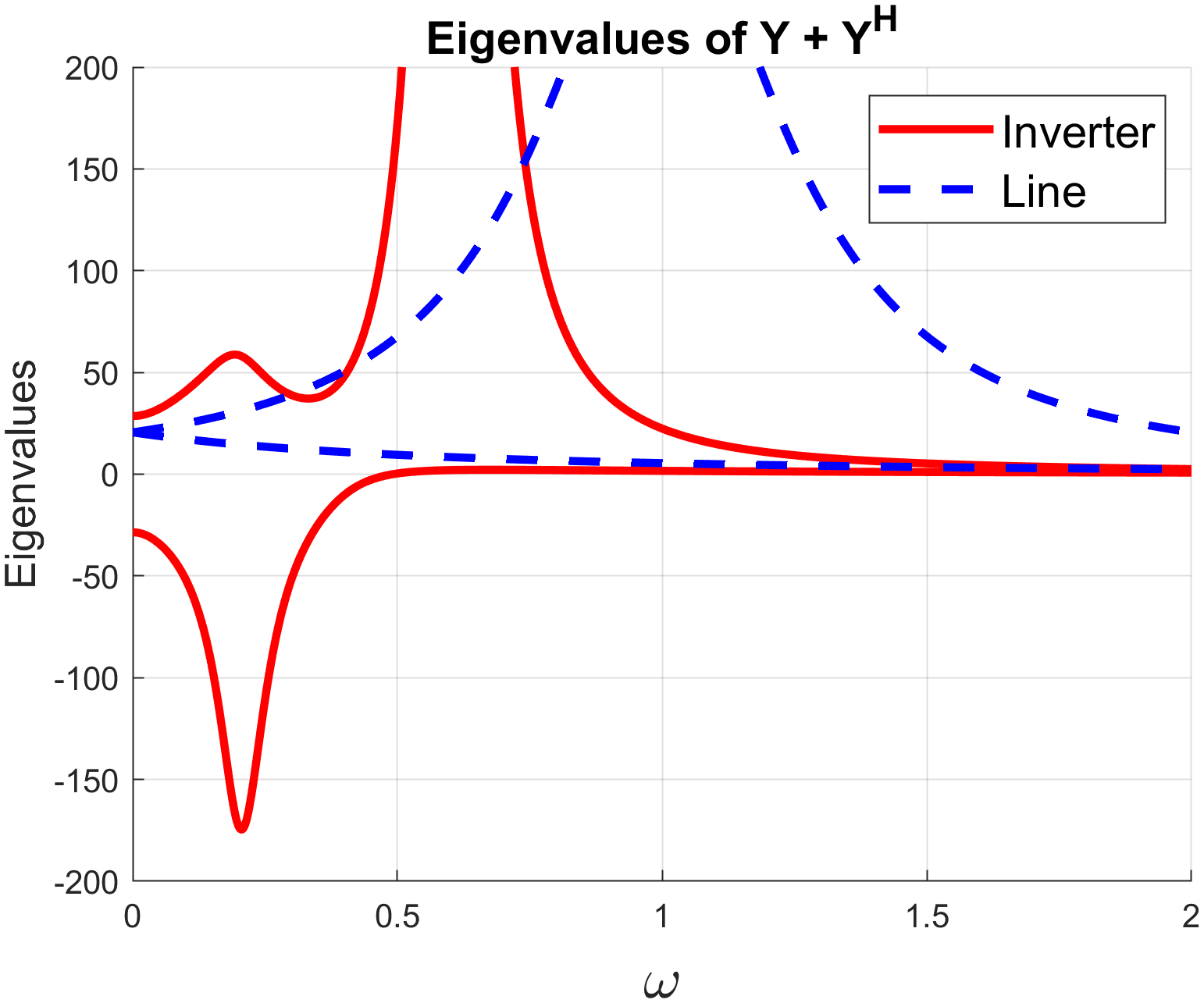}
    \caption{Eigenvalues of the Hermitian part of the admittance matrix of a GFM  - red curve, and an $R-L$ line - dashed blue curve.  }
    \label{fig:one_inverter}
    
\end{figure}

As an example of a non-passive device, Figure \ref{fig:one_inverter} shows the eigenvalues of the $\bf{Y}_{D,ii}(j\omega) + \bf{Y}_{D,ii}(j\omega)^H$ matrix for a droop-controlled GFM. The GFM model is taken from \cite{pogaku2007modeling} and represents a $13$th order model that includes the explicit modeling of coupling impedance, $LCL$-filter, current control loop, voltage control loop, and real and reactive power controllers. While both eigenvalues corresponding to $R-L$ line admittance are always positive since $R-L$ line is a passive element, one of the eigenvalues for the GFM inverter becomes negative at low frequencies. Hence, the corresponding matrix (i.e., Hermitian part of the admittance matrix) becomes sign indefinite at low frequencies, and this can not be corrected by simply adjusting control parameters - whatever the values of the voltage and frequency droop coefficients one chooses, there is is always one of the eigenvalues becoming negative at low enough frequencies.

In order to modify the stability conditions \eqref{pos_def_main} for practical use, we return back to equation \eqref{det_freq} and note that it remains valid if the system admittance matrix $\bf{Y}_F$ is multiplied by another arbitrary non-singular matrix $\bf{M}$. After this multiplication, instead of \eqref{pos_def_main} we have a more general condition:
\begin{equation}\label{pos_def_mod}
   \bf{M}(\omega)\bf{Y_F}(j\omega,\alpha) + \bf{Y_F}^H(j\omega,\alpha)\bf{M}^H(\omega)  \succ 0
\end{equation}
We note here, that the only essential property of the multiplier matrix $\bf{M}$ is a non-singularity, but $\bf{M}$ does not have to be positive or negative definite. This is in contrast to the requirement imposed on \emph{passivity multipliers} that have to be represented by stable transfer-functions \cite{desoer2009feedback}. We also note, that the matrix $\bf{M}$ can be a function of the homotopy parameter $\alpha$ - for the sake of simplicity we will not explicitly indicate this in the equations. 
On the other hand, our method requires condition \eqref{det_freq} to stay valid all the way along the homotopy path. Condition \eqref{pos_def_mod} is similar to a specific realization of the method of \emph{integral quadratic constraints} (IQC), however, unlike in the original paper \cite{megretski1997system} the multiplier is applied to the linear plant instead of scaling the feedback.

From \eqref{pos_def_mod} we now can formulate the fully decentralized stability conditions. To do this, we choose the multiplier matrix $\bf{M}$ to be block-diagonal, with blocks represented by $2$ by $2$ matrices:
\begin{equation}\label{eq:big_M}
\bf{M}(\omega) =
\begin{bmatrix}
\bf{m}_1(\omega) & 0 & \cdots & 0 \\
0 & \bf{m}_2(\omega) & \cdots & 0 \\
\vdots & \vdots & \ddots & \vdots \\
0 & 0 & \cdots & \bf{m}_i(\omega)
\end{bmatrix}
\end{equation}
Such a choice results in the diagonal elements of the device admittance matrix  $\bf{Y}_F$ being multiplied by the corresponding blocks $\bf{m}_i$: 
\begin{equation}\label{m_multi}
    (\bf{M}\bf{Y_{D}})_{ii} = \bf{m}_i\bf{Y_{D,ii}}
\end{equation}
Therefore, instead of conditions \eqref{pass_network} and \eqref{pass_devs} we now have more general conditions: 
\begin{equation}\label{main_lines}
    \bf{M}(\omega)\bf{Y}_{N}(j\omega, \alpha) + \bf{Y}_{N}^H(j\omega,\alpha)\bf{M}^H(\omega) \succ 0 
\end{equation}
\begin{equation}\label{main_devs}
    \forall i, \quad \bf{m}_i(\omega)\bf{Y}_{D,ii}(j\omega,\alpha) + \bf{Y}_{D,ii}^H(j\omega,\alpha)\bf{m}_i^H(\omega) \succ 0 
\end{equation}
Thus, we need to find such set of $2$ by $2$ matrices $\bf{m}_i$ (if it exists) that simultaneously ensures both \eqref{main_lines} and \eqref{main_devs}. The condition for the matrix $\bf{M}$ in \eqref{eq:big_M} being non-singular is equivalent to demanding for each block $\bf{m}_i$ to be non-singular. Apart from non-singularity, there are no other restrictions for blocks $\bf{m}_i$ - they do not have to have any special structure or properties (such as being Hermitian/symmetric or real valued etc.). We note, that $\bf{m}_i$ can be functions of frequency and, in particular, can be chosen to be piece-wise constant. We also note, that the network admittance matrix $\bf{Y}_N$ having elements of the form \eqref{Y_line} automatically takes into account the network electro-magnetic dynamics, thus making our method naturally fit for power electronics dominated grids. 

Conditions \eqref{main_lines} and \eqref{main_devs}, if satisfied by a certain choice of the set of $\bf{m}_i$, certifies stability for a given set of devices connected to the network with the admittance matrix $\bf{Y}_N$ in an arbitrary way - i.e., any device from the given set can be connected to any bus of the network without violating the stability. This property can be referred to as \emph{plug-and-play} capability. 

An even more flexible certificates can be obtained by setting all the $2$ by $2$ blocks in \eqref{eq:big_M} to be the same, i.e.,  $\bf{m_i}=\bf{m}$. In this case the result of the multiplication of the network matrix $\bf{Y}_N$ by $\bf{M}$ can be conveniently reduced to individual lines: 
\begin{equation}\label{m_multi}
    (\bf{M}\bf{Y_{N}})_{ik} = \bf{m}\bf{Y_{N,ik}}
\end{equation}
and the conditions \eqref{main_lines} and \eqref{main_devs} can be recast as a set of independent conditions for every line and device type: 
\begin{equation}\label{ind_lines}
    \forall i,k,  \quad \bf{m}(\omega)\bf{Y}_{N,ik}(j\omega) + \bf{Y}_{N,ik}^H(j\omega)\bf{m}^H(\omega) \succ 0 
\end{equation}
\begin{equation}\label{ind_devs}
    \forall i,  \qquad \bf{m}(\omega)\bf{Y}_{D,ii}(j\omega) + \bf{Y}_{D,ii}^H(j\omega)\bf{m}^H(\omega) \succ 0 
\end{equation}
In this case, there is a complete \emph{plug-and-play} compatibility - the given set of devices can be connected with the given set of lines in a fully arbitrary way - the network topology does not matter - every configuration will be stable.

We note here, that only the type of line - i.e., its $R/X$ ratio is important for \eqref{ind_lines}, not the length of the line since the latter only multiplies the admittance matrix by a factor, and does not influence the positive definiteness. We also note, that conditions \eqref{ind_lines} do not have to be checked for every type of line in the network - it is sufficient to check the lines with the maximum and minimum $R/X$ ratio - the proof can be found in the Appendix \ref{prop:extreme_rx}. Note that this property is important from a vendor perspective as a vendor wants to develop IBR controller which could be connected to any system. Our stability certificates mean that if stability is certified for a device and transmission lines within certain $R/X$ ratio range, the device can be connected to any network with any topology as long as the lines are within that $R/X$ ratio range.

Note that the matrix $\bf{m}$ can be piece-wise constant with respect to frequency. Hence, the whole frequency range can be divided into segments, with different $\bf{m}$ for each segment, hence allowing one to check the stability conditions \eqref{ind_lines} and \eqref{ind_devs} for each segment separately. This is important as the attention can be concentrated on finding such $\bf{m}$ (if it exists) that satisfies \eqref{ind_lines} and \eqref{ind_devs} in a frequency range that did not satisfy the original passivity conditions \eqref{pass_network} and \eqref{pass_devices}. In \cite{vorobev2019decentralized} we have demonstrated that property using a simple heuristic choice of $\bf{m}$ with a different $\bf{m}$ for a low frequency range, which was  originally non-passive, than a high-frequency range which was passive:
 \begin{equation}\label{m_simple}
\bf{m}(\omega) =
\begin{cases}
\begin{bmatrix}
0 & -1 \\
1 & 0
\end{bmatrix}, & \omega \leq \Omega_0, \\[1em]
\begin{bmatrix}
1 & 0 \\
0 & 1
\end{bmatrix}, & \omega > \Omega_0.
\end{cases}
\end{equation}
where $\Omega_0 = 2\pi \cdot 50$ rads/s. That choice of $\bf{m}$ ensured that the eigenvalues of the resulting matrix in \eqref{ind_devs} were positive in the whole frequency range. Note that at high frequencies the positive definiteness is already satisfied, as is seen in Figure \ref{fig:one_inverter}, so $\bf{m}$ is the unity matrix.

\begin{figure}
    \centering
    \includegraphics[width=0.7\linewidth]{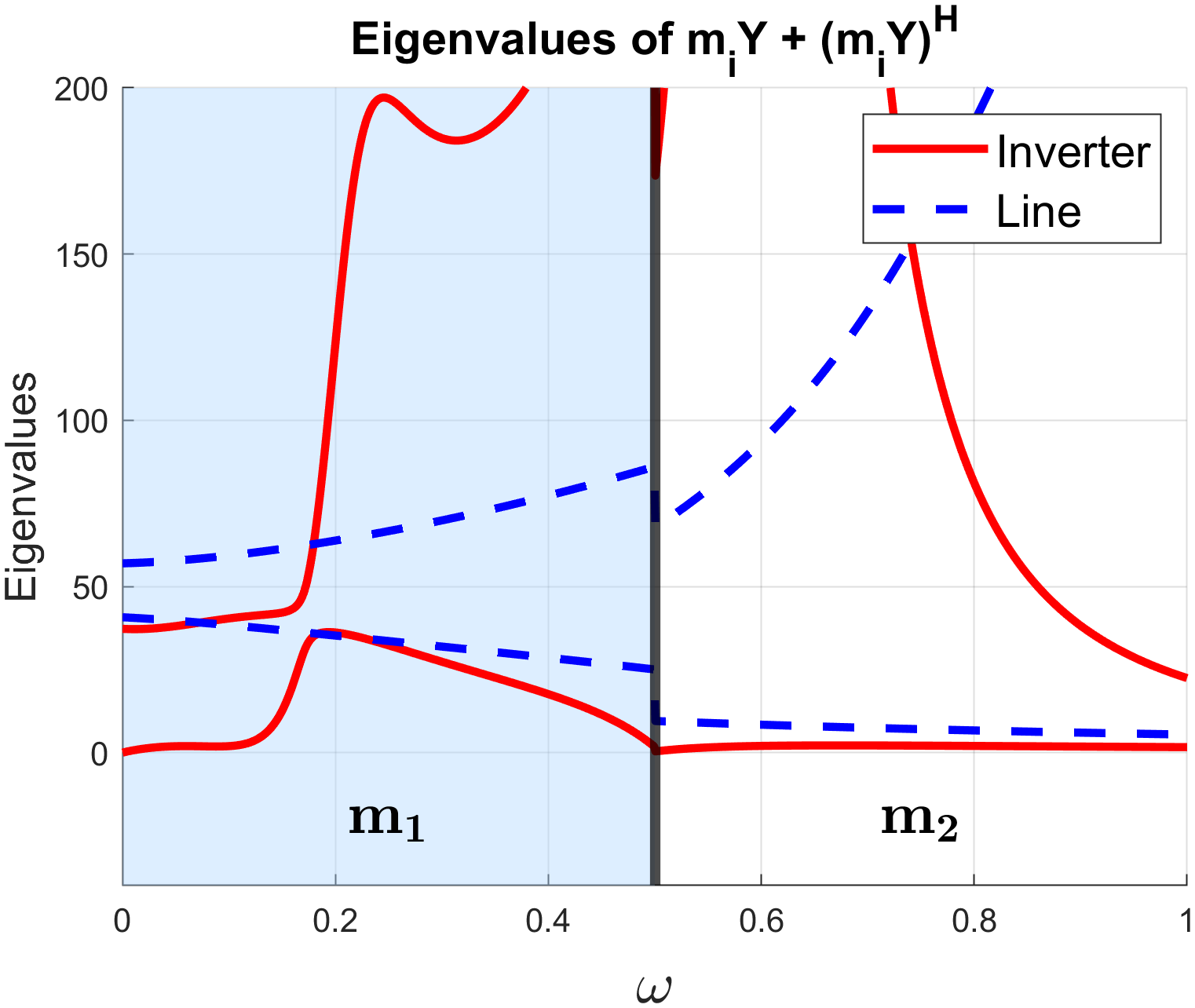}
    \caption{ Transformation of the inverter and line admittance matrices by a piecewise matrix with $\bf{m}_1$ and $\bf{m}_2$ from \eqref{m_SDPe} in two regions of frequency. Frequency axis is normalized by the nominal power grid frequency.  }
    \label{fig:two_regoins_split}
\end{figure}

The above choice of the transformation matrix $\bf{m}$ was heuristic and resulted in conservative stability certificates, but one can search for a more systematic choice of $\bf{m}$ that leads to conditions
\eqref{ind_lines} and \eqref{ind_devs} to be simultaneously satisfied. In fact, the problem can be cast as a semi-definite programming with constraints (SDP), which is a convex setup \cite{boyd2004convex}.
In practice, the constant multiplier $\bf{m}$ can be synthesized by solving a semidefinite program on a discretized frequency grid (here, 100 samples per frequency interval), maximizing the sum of minimum eigenvalues for every frequency sample $\omega_k$ subject to
\begin{equation}
\begin{aligned}
\max_{\mathbf{m},\{t_k\}} \quad & \sum_{k=1}^{N_\omega} t_k \\
\text{s.t.} \quad 
& \mathbf{m}\mathbf{Y}(j\omega_k)+\mathbf{Y}^\mathbf{H}(j\omega_k)\mathbf{m}^\mathbf{H} \succeq t_k I,
\end{aligned}
\label{eq:sdp_m}
\end{equation}
where each scalar $t_k$ is the smallest eigenvalue of $(\mathbf{m}\mathbf{Y}+\mathbf{Y}^\mathbf{H}\mathbf{m}^\mathbf{H})$. In this way, the SDP directly searches for a multiplier that makes the Hermitian part positive over the whole sampled frequency range, consistent with the multiplier-based certificate used here. An alternative approach is to synthesize a \emph{dynamic} multiplier $\mathbf{m}(s)$ by parameterizing it as a stable LTI filter as we show in \cite{gorbunov2026dynamic}. Detailed description of the SDP problem formulation and solution procedure used for the division of the frequency range into two and three segments will be reported in a separate publication. For instance, for the same system with droop-controlled grid-forming inverters and similar split of the frequency into two regions: $\omega < 0.5\Omega_0$ and $\omega>0.5\Omega_0$, one can find using SDP the following two matrices for the corresponding regions:
\begin{subequations}\label{m_SDPe}
\begin{equation}
\mathbf{m}_1 = 
\begin{bmatrix}
0.204 - 0.048 j & -0.652 - 0.106 j \\
0.607 - 0.381 j & 0
\end{bmatrix} 
\end{equation}
\begin{equation}
\mathbf{m}_2= 
\begin{bmatrix}
1 & 0 \\
0 & 1
\end{bmatrix}.
\end{equation}
\end{subequations}

Figure \ref{fig:two_regoins_split} shows the graphs of eigenvalues for both inverter and line for this case and proves that both sets of eigenvalues are turned positive in every region. 
          
One can further reduce conservativeness of the methodology by splitting frequency into three segments and using SDP-based method to find optimal values of the matrices. The frequency regions are: $\omega <  0.4 \Omega_0$, $0.4 \Omega_0 < \omega < 0.6\Omega_0$, and $0.6\Omega_0 < \omega$. The values for specific matrices $\mathbf{m}_1$, $\mathbf{m}_2$, and $\mathbf{m}_3$ used in each of these regions are:  
\begin{subequations}\label{m_SDP_3}
\begin{equation}
\mathbf{m}_1 = 
\begin{bmatrix}
0.145 - 0.048j & -0.637 - 0.194j \\ 
0.524 - 0.506j &  0.000 - 0.000j 
\end{bmatrix}, 
\end{equation}
\begin{equation}
\mathbf{m}_2 = 
\begin{bmatrix}
0.065 + 0.467j &  -0.57 + 0.031j \\
0.478 - 0.007j &  0.065 + 0.469j 
\end{bmatrix}, 
\end{equation}
\begin{equation}
\mathbf{m}_3 = 
\begin{bmatrix}
1 & 0 \\
0 & 1
\end{bmatrix}.
\end{equation}
\end{subequations}

\begin{figure}
    \centering
    \includegraphics[width=0.7\linewidth]{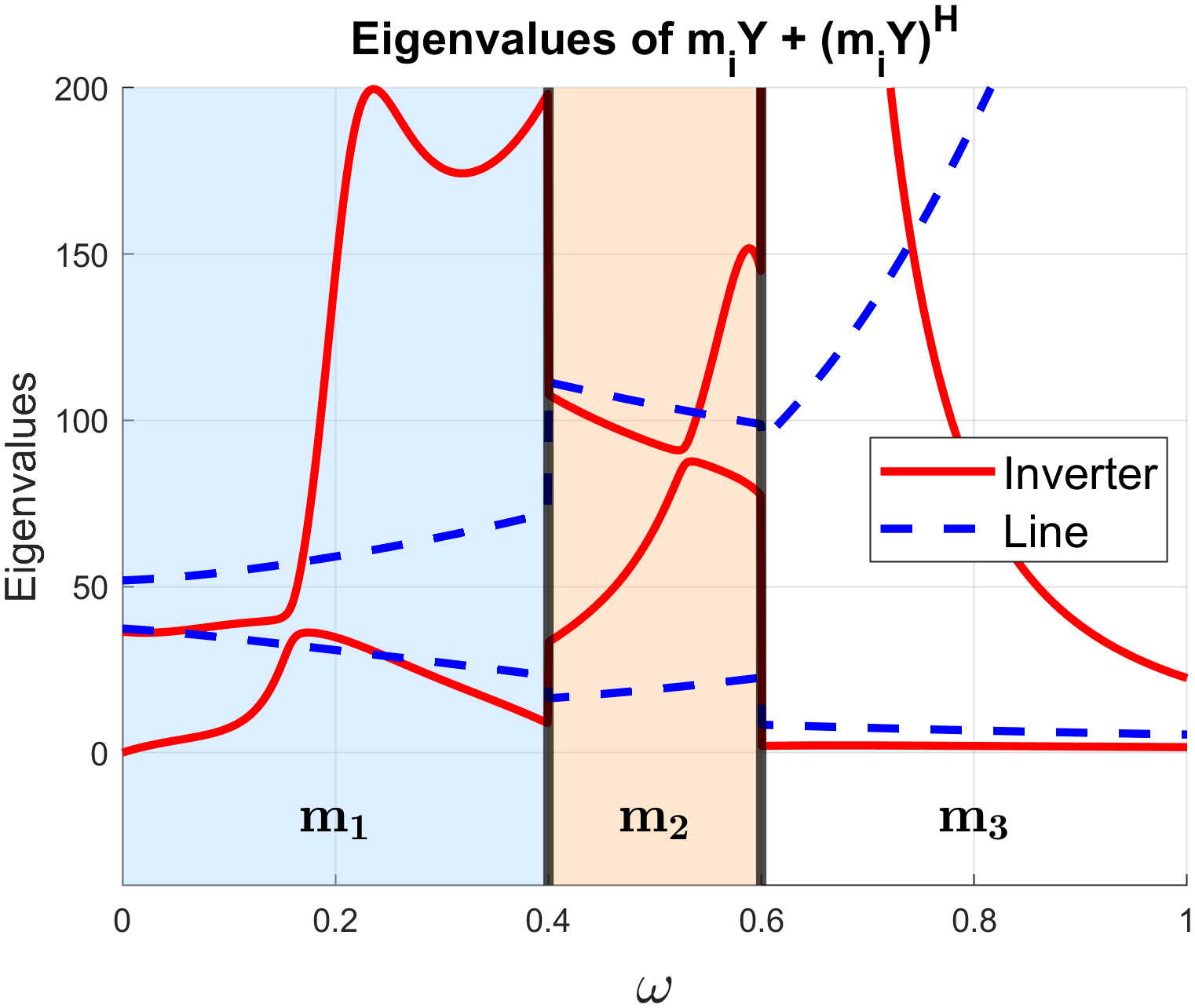}
    \caption{ Transformation of the inverter and the line admittance matrices by a piecewise matrix with $\bf{m}_1$, $\bf{m}_2$, and $\bf{m}_3$ from \eqref{m_SDP_3} for three regions of frequency.  }
    \label{fig:three_regoins_split}
\end{figure}

The corresponding eigenvalues for the transformed admittance matrices of the inverter and the line are shown in Fig. \ref{fig:three_regoins_split} and prove that both sets of eigenvalues are turned positive in every region.

Individual admittance matrices of the line elements in \eqref{main_lines} depend only on the line inductance and resistance, and not on the operating point. Device admittance matrices in \eqref{main_devs} on the contrary, are the functions of the operating point. The role of the operating point for every device is two-fold: the varying terminal voltage at the operating point, and the varying device settings (eg., real and reactive power output). This means that the certificates depend on the operating point and that problem is the subject of further research.

\section{Black-box model approach}

One of the main advantages of our method is its natural generalization to situations when one does not have access to explicit "white-box" type of models for the devices. Manufacturers regard their control algorithms as critical
proprietary technology and tend to disclose to system operators
only black-box EMT models which can be plugged into the rest of the grid EMT model to carry
out time-domain simulation. But the black-box IBR models can also be used to generate input-output impedance characteristics in the frequency domain. Alternatively, the impedance characteristic of an IBR can be obtained in a lab using input signals of different frequencies. Either way gives us the required admittance matrix over the whole frequency range, which we can use to apply our method to certify plug-and-play stability.  

\begin{figure}
    \centering
    \includegraphics[width=0.7\linewidth]{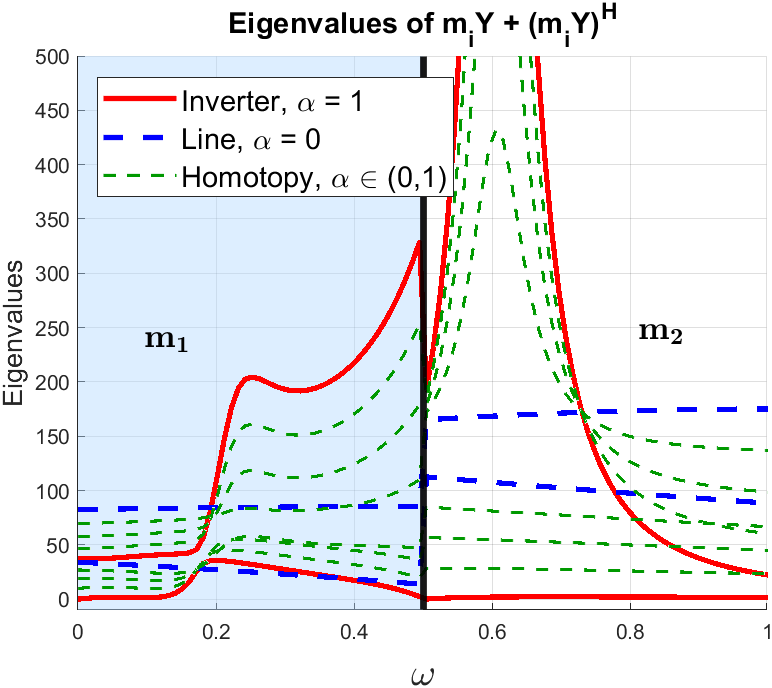}
    \caption{Homotopy from a passive R-L element to a GFM inverter as described by equation \eqref{data_driven_homotopy}.
   }
    \label{fig:homotopy_data_driven}
\end{figure}

As is seen from conditions \eqref{main_devs} or \eqref{ind_devs} for devices, only the admittance matrix for all frequencies is needed, not the explicit model of a device. The only challenge here is that we need to perform a homotopy from a definitely stable state and make sure that conditions \eqref{main_devs} or \eqref{ind_devs} are valid all the way through the homotopy. In the previous examples in this paper we have set controller gains to zero to obtain a stable starting configuration for the homotopy but it is not possible to do so when we use experimentally obtained admittance spectra. There are many ways one can perform the homotopy in that case, and the choice of proper homotopy path for a given type of device is a separate research question. Here we note, that for devices that do not have right-hand zeros (so called, transmission zeros for multiple-input multiple-output systems), one can use a very simple homotopy starting with a certain (fictitious) passive element (say, $R-L$ element) instead of the device in question, and then gradually transforming the said $R-L$ element into the device, i.e:
\begin{equation}\label{data_driven_homotopy}
    \bf{Y}_{hom} (\alpha,j\omega) = \alpha \bf{Y}_{D,ii}(j\omega) + (1-\alpha) Y_{pass}(j\omega) 
\end{equation}
here $\bf{Y_{hom}}$ is the admittance along the homotopy path, and $\bf{Y_{pass}(j\omega)}$ is the admittance of the chosen fictitious passive element. The choice of parametrization \eqref{data_driven_homotopy} is not unique, and can be done depending on the specific configuration either to simplify the calculations or to provide less conservative results. Figure \ref{fig:homotopy_data_driven} shows the eigenvalues for a series of admittance matrices of equation \eqref{data_driven_homotopy} (already modified by matrices $\bf{m}$ from \eqref{m_simple} as the homotopy parameter $\alpha$ varies from $0$ to $1$. The dashed blue line shows the eigenvalues for the case when $\alpha=0$ - i.e., when inverter is replaced by a passive element, while the solid red lines correspond to $\alpha=1$ - when passive element is "turned off", and only the inverter remaining. A series of dashed green lines correspond to intermediate values of the homotopy parameter value $\alpha$ in \eqref{data_driven_homotopy} - we see, that all the eigenvalues remain positive as the homotopy is performed. Thus, we demonstrate, that in order to certify the plug-and-play stability of a system of droop-controlled inverters we do not in fact need their explicit models, just the actual admittance values at given operating conditions. That admittance spectrum can be obtained by performing an impedance scan in a lab using an actual device or from a black-box model (compliled EMT simulation code) provided by a vendor.

\section{Validation}

\begin{figure}
    \centering
    \includegraphics[width=0.65\linewidth]{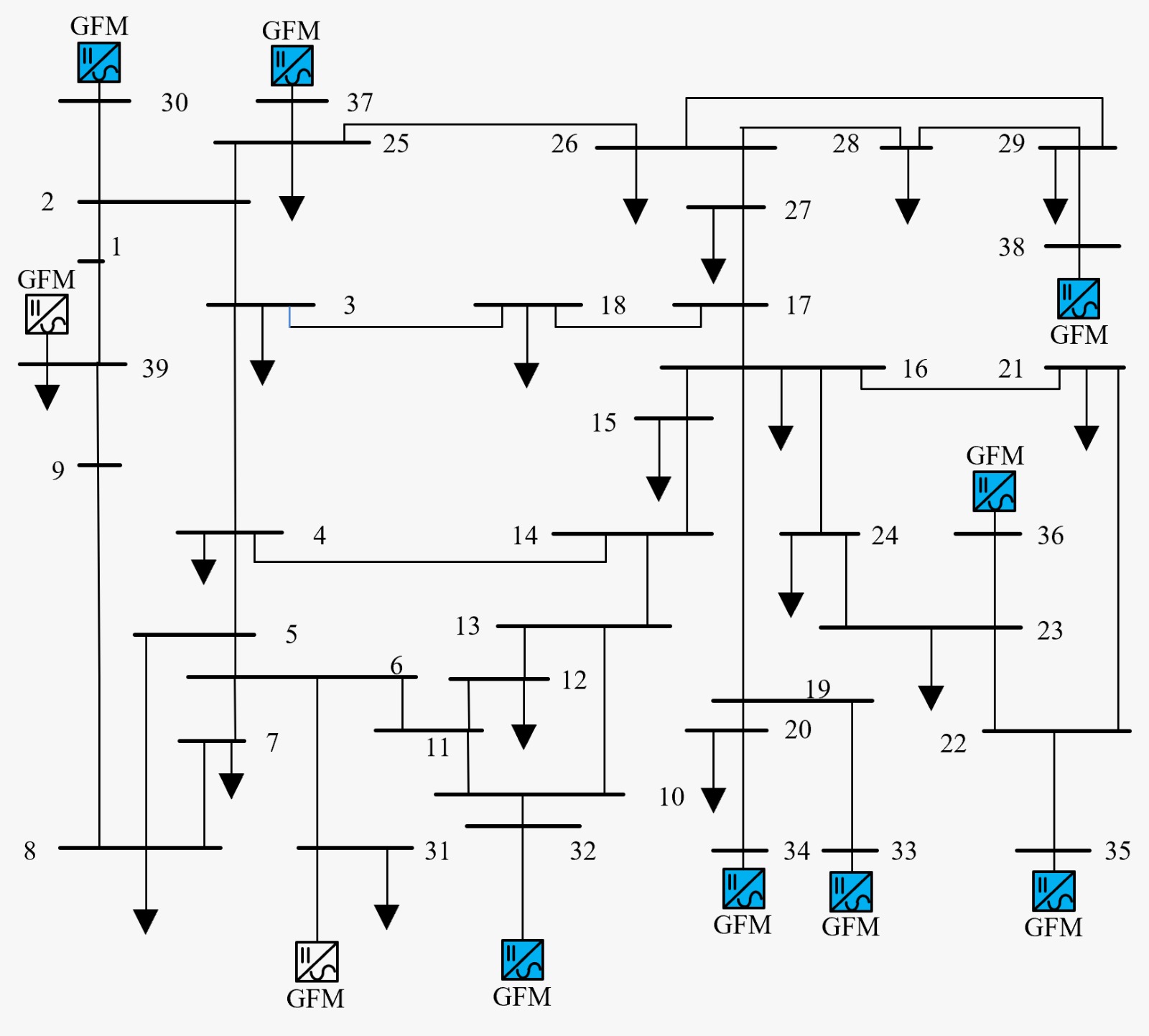}
    \caption{IEEE 39-bus test case modified for testing stability for IBR dominated grids: every generator in the system is replaced by a grid-forming inverter.}
    \label{fig:IEEE39_case}
\end{figure}

In this section we provide validation of our proposed method on a realistic test-case for a power system. We use the well-known IEEE 39-bus test case which in its original form contains 10 synchronous generators. For our testing we substitute all the generators with GFMs (GFM model is taken from \cite{pogaku2007modeling}) - a possible future scenario for 100\% renewable grids (Fig. \ref{fig:IEEE39_case}). Each GFM is rated at $200$ MVARs and the loading level for the system is scaled down correspondingly. Since our goal is to demonstrate the plug-and-play capability of our method, we adopt the following strategy for our validation. We test a large number of scenarios with different placement of these $10$ GFMs in the grid. Since the total number of all possible combinations for GFMs placement across the grid is way too big - more than $20$ million possible combinations for the selected test-case, we randomly choose some limited number of combinations - $200$ combinations - and perform direct stability assessment of each chosen case by calculating the eigenvalues. For simplicity the leakage inductance of all transformers is set to the same value of $0.025$ pu on the basis of $200$ MVAR and $345$ kV for power and voltage respectively, and the resistance of every transformer is set to $0.0025$ pu.     

\begin{figure}
    \centering
    \includegraphics[width=0.7\linewidth]{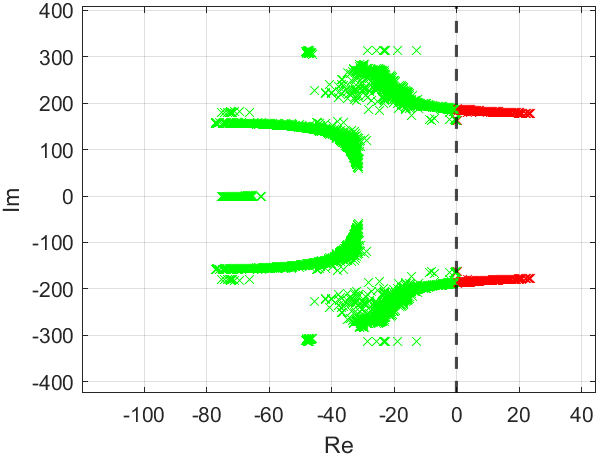}
    \caption{Eigenvalues (only dominant ones shown) for $200$ scenarios with different placement of $10$ inverters in the IEEE $39$-bus test case. }
    \label{fig:base39_200}
\end{figure}

We first set the frequency droop coefficient $k_p$ of every GFM to be $5\%$ and the voltage droop coefficient $k_q$ to be $1\%$. We now test $200$ randomly chosen GFM placement combinations using these settings. We find, that there are both stable and unstable combinations present. This could be expected, since stability is dependent on the actual system configuration, not only on the settings of individual devices - out of $200$ generated scenarios $8$ are stable and $192$ are unstable. Figure \ref{fig:base39_200} shows the eigenvalues of all those $200$ test cases (only the dominant poles are shown) - it is clear that  a rather big number of modes are located in the right half-plane. It is known that for GFMs in general, stability is enhanced by reducing the voltage and frequency droop coefficients. If we reduce the frequency droop of every inverter to $4\%$ (keeping voltage droop the same as before - $1\%$) we still get quite a big number of unstable cases - from the new set of randomly generated $200$ cases we now have $164$ unstable and $36$ stable ones.

The testing results of the previous paragraph suggests that: a) for a realistic case one can not guarantee stability by considering some number of specific configurations - the system can still be non plug-and-play compatible even if we managed to generate some stable cases; b) when we move closer to the true plug-and-play stability boundary (if it exists for the system in question), it becomes harder to locate possible unstable configurations, and a false positive decision for stability can be made based on direct assessment of some number of configurations.

Next, we demonstrate the use of our method by finding the allowed region for the droop coefficients values that will guarantee plug-and-play stability. We keep the transformer leakage reactance and resistance of every GFM $0.025$ and $0.0025$ pu respectively. Then, using our method and a full inverter model from \cite{pogaku2007modeling} we construct a region in the $k_p$-$k_q$ plane for every inverter that guarantees plug-and-play stability for the whole system. \footnote{It is not expected that System Operator would calculate the stability region as it would require a very significant computational effort- here it is just a test to check conservativeness of the method} For this we use three different choice of transformation matrices $\bf{m}$ in \eqref{ind_lines} and \eqref{ind_devs} - piecewise constant matrices for each set: 1) split of frequency range into two parts, two heuristically-chosen $\bf{m}$ matrices  \eqref{m_simple} taken from  \cite{vorobev2019decentralized}, 2) split of frequency range into two parts, two $m$ matrices from \eqref{m_SDPe}, obtained using SDP approach, and 3) split of frequency range into three parts, three $\bf{m}$ matrices in \eqref{m_SDP_3}, obtained using SDP approach.

\begin{figure}
    \centering
    \includegraphics[width=0.9\linewidth]{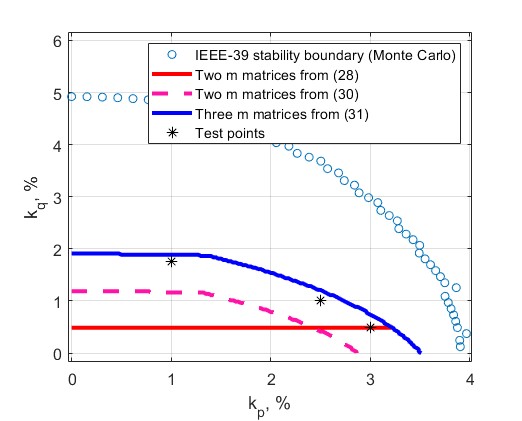}
    \caption{Plug-and-play stability regions in the space of frequency and voltage droop of a GFM obtained using different sets of transformation matrices $\bf{m}$. }
    \label{fig:regions_3}
\end{figure}

Figure \ref{fig:regions_3} shows plug-and-play stability regions obtained using our proposed method for the above listed three different choices of transformations $\bf{m}$. A heuristic choice from \eqref{m_simple} can already provide some stability certification - a solid red line in Fig. \ref{fig:regions_3}, although the certified region in $k_p$-$k_q$ space is rather narrow. A bigger region can be obtained if we perform a search for the transformation matrix $\bf{m}$ at low frequencies using an SDP approach, see \eqref{m_SDPe}, and the corresponding stability region is shown with a dashed magenta line in Fig. \ref{fig:regions_3} - this region is already quite reasonable for practical implementations. Finally, we try to certify stability by searching for three matrices $\bf{m}$, i.e., splitting the frequency range into three pieces. The resulting matrices from \eqref{m_SDP_3} provide the stability region shown by solid dark blue curve on Fig. \ref{fig:regions_3} - this region is considerably bigger than the previous two. It is expected that dividing the frequency spectrum into a larger number of segments with different transformation matrices would further reduce conservativeness of the method, which is the subject of our further research.

By definition, plug-and-play stability region means that we can arbitrarily choose the droop coefficient pairs for any inverter from this region, and stability of the system with arbitrary interconnection of arbitrary number of such inverters is guaranteed. Let us now test our certificates from Fig. \ref{fig:regions_3}. For this we choose three pairs of frequency and voltage droops ($k_p$,$k_q$), namely ($3\%$, $0.5\%$), ($2.5\%$, $1\%$), and ($1\%$, $1.75\%$) - all from within the region bounded by the dark blue solid curve, with each point lying close to the region boundary (shown by asterisks in Fig. \ref{fig:regions_3}). For each of the chosen droop pair we run $200$ scenarios of random placement of $10$ inverters with each inverter having the same values of droop coefficients. For each three cases this results in all the scenarios being stable. Finally, we choose to assign different droop coefficients to inverters in the system - $5$ inverters having ($k_p$,$k_q$) equal to ($2.5\%$, $1\%$) and the other $5$ inverters - to ($1\%$, $1.75\%$). We again run $200$ randomly generated inverter placement scenarios that all show to be stable.

Next, we attempt to assess the conservativeness of our method: in addition to secure plug-and-play stability regions, Fig. \ref{fig:regions_3} also shows the numerically estimated region beyond which plug-and-play stability is lost - region bounded by blue circles. Each circle corresponds to the case when we fix the ratio of voltage and frequency droop and increase the gain values gradually until we find at least one unstable case out of the $200$ scenarios being probed. Such a method provides an upper boundary - plug-and-play stability is not expected to be beyond the region bounded be blue circles. In the region between the solid red curve and the blue circles there could be some unstable scenarios which are simply not covered by our random selection for direct simulation. However, the true plug-and-play stability boundary is not possible to obtain by direct simulations - one will need to check all the possible combinations for inverters placements. The number of this combinations is of the order of millions to tens of millions, which makes it infeasible to perform direct numerical analysis. This means that the distance between the red curve and the blue circle curve in Fig. \ref{fig:regions_3} gives only an upper-bound assessment of conservativeness of the methodology as the actual distance may be smaller - but is not possible to calculate precisely.

\section{Practical application of the method}

In this section we present a possible practical application of the proposed method for real-life power systems with IBRs. We assume that System Operator has black-box models of all the IBRs. According to equations \eqref{ind_devs} the task of System Operator is to find such a $2$ by $2$ matrix $\bf{m}$ that makes the Hermitian part of the admittance  matrix of every device (and every line) in the system positive definite. At first, this looks like an infeasible problem since the number of devices in the grid is very large. However, in reality it is sufficient to demonstrate the positive definiteness for every \emph{device type}. The number of types of devices in the grid can be rather modest, in particular if certain standardization codes are implemented. Moreover, for any vendor it is economically advantageous to mass-produce devices of the same type, rather than having many types. Thus if the grid contains, say, $10000$ devices belonging to $10$ different types, we need to find such a transformation matrix $\bf{m}$, that makes all $10$ Hermitian parts positive definite - a very much computationally feasible problem. Note that the computational complexity of the method does not depend on the system size but only on the number of types of devices.

Let us now assume that System Operator has managed to find such a matrix $\bf{m}$ that makes the Hermitian part of the admittance matrix of all the device types positive definite (and also all the lines in the grid). In this case these devices can be connected to the grid arbitrarily and stability will be certified - very much like we demonstrated in the previous section for the set of GFMs. Let us now outline a procedure for connecting a new device to this grid. First, if a new device is of the same type as one of those previously assessed, we can connect the device without any additional evaluation and, what is important, we do not need to modify the matrix $\bf{m}$. If the type of device is new, the vendor needs to check if the Hermitian part of the admittance matrix for the device is positive definite under the known transformation matrix $\bf{m}$: 
\begin{itemize}
\item 
If the answer is "yes" - we can safely connect the device to the grid. We do not need to re-evaluate all other models, and don't even have to know all other models - just the common matrix $\bf{m}$. This is very convenient from the privacy point of view - the grid operator discloses to a new vendor only the common $\bf{m}$ matrix, not the models of other devices. 
\item If the answer is "no", two options are possible: 
\begin{enumerate}
    \item The grid operator can do the full re-evaluation trying to assess whether a new matrix $\bf{m}$ exist that works on the full set of device types - the existing ones and the newly connected one
    \item The vendor of the new device type can adjust the device controls so that it becomes compliant with the existing matrix $\bf{m}$ and therefore can be safely connected to the grid.
\end{enumerate}

\end{itemize}

The above listed procedure offers a rather practical way for adding new device types to the grid if there already exist a common stability certification framework - mainly the transformation matrix $\bf{m}$ is known for the devices already in operation. Finding such matrix $\bf{m}$ that is suitable for certain set of device types without imposing unreasonably tight constraints is a separate procedure and a subject of further research. We envision that with the increased IBR share the question of \emph{equipment standardization} will become necessary, and industrial standards can also be developed for transformation matrices $\bf{m}$. In this case any vendor will need to make sure that their devices are compliant with this $\bf{m}$ before commissioning them - this is somewhat similar to the requirement of electromagnetic compatibility that is routinely checked at present.

\section{Conclusions}

In this paper we have presented an original method for deriving decentralized, plug-and-play stability certificates which could be obtained using either black-box admittance spectra, or white-box models of devices (if these are available). The method is based on analyzing the properties of the nodal admittance matrix itself, without recasting the problem as a feedback interconnection between a device and a known external network. Although we tested our method directly on a system with grid-forming inverters, the mathematical foundation of the method allows its application to any type of device. Details of such applications and building of specific approaches is a subject of further research.

Although the method is related to the classical passivity concept (with generalizations) but it is derived independently, without relying on any prior results from passivity theory. The method introduces frequency-dependent transformation matrices $\bf{m}$, enabling stability certification when controllers fail strict passivity tests in a certain frequency range. The certificates can be made topology-independent: a certified device can be connected to any node of a network with any topology, provided transmission lines fall within a specified $R/X$ ratio range. We demonstrate that the method has rather modest conservativeness, which can be further reduced by dividing the frequency range into more segments with optimized transformation matrices for every segment, as demonstrated on the IEEE $39$-bus test case. Computational complexity of the methodology depends only on the number of device types, but not the network size, making the method scalable to large-scale systems.

\appendices

\appendix[Proof of the linearity of the line admittance with respect to R/X ratio] \label{prop:extreme_rx}

For RL line admittance $\bf{Y}_{N,ik}(\rho)$ parameterized by $\rho=R/X$, if $\bf{m}\bf{Y}_{N,ik}(\rho) + \bf{Y}_{N,ik}(\rho)^H m^H \succ 0$ holds at $\rho_{\min}$ and $\rho_{\max}$, then it holds for all $\rho\in[\rho_{\min},\rho_{\max}]$.

\emph{Step 1}: For any nonsingular $A$, 
\begin{equation}
 A + A^H \succ 0 \Leftrightarrow  A^{-1} + (A^{-1})^H \succ 0.
\end{equation}

Since $Z(\rho)=Y(\rho)^{-1}$ is linear in $\rho$, the Hermitian part 
\begin{equation} \label{eq:Hermitian_of_Z}
Z(\rho)m^{-1} + (m^{-1})^H Z(\rho)^H
\end{equation}
is also linear in $\rho$. By Step 1 above and by convexity of positive definiteness, if this holds at both endpoints, it holds throughout the interval.

\ifCLASSOPTIONcaptionsoff
  \newpage
\fi

\bibliographystyle{IEEEtran}
\bibliography{bibtex/bib/bibliography}

\end{document}